\documentclass[twocolumn,showpacs,superscriptaddress,english,aps,prl]{revtex4}
\usepackage[T1]{fontenc}
\usepackage[latin9]{inputenc}
\usepackage{amstext}
\usepackage{graphicx}
\usepackage{amssymb}
\usepackage{bm}

\makeatletter

\providecommand{\LyX}{L\kern-.1667em\lower.25em\hbox{Y}\kern-.125emX\@}
\newcommand{\lyxmathsym}[1]{\ifmmode\begingroup\def\b@ld{bold}
  \text{\ifx\math@version\b@ld\bfseries\fi#1}\endgroup\else#1\fi}

\@ifundefined{textcolor}{}
{%
 \definecolor{BLACK}{gray}{0}
 \definecolor{WHITE}{gray}{1}
 \definecolor{RED}{rgb}{1,0,0}
 \definecolor{GREEN}{rgb}{0,1,0}
 \definecolor{BLUE}{rgb}{0,0,1}
 \definecolor{CYAN}{cmyk}{1,0,0,0}
 \definecolor{MAGENTA}{cmyk}{0,1,0,0}
 \definecolor{YELLOW}{cmyk}{0,0,1,0}
 }

\makeatother

\usepackage{babel}

\begin{document}

\title{Metal to insulator transition on the N = 0 Landau level in graphene}

\author{Liyuan Zhang}

\affiliation{CMPMSD, 
 Brookhaven National Laboratory, Upton, NY 11973 USA}

\author{Yan Zhang}
\affiliation{Department of Physics and Astronomy, Stony Brook University, Stony Brook, New York 11794-3800, USA}
\affiliation{Center for Functional Nanomaterials,
 Brookhaven National Laboratory, Upton, NY 11973 USA}

\author{M.~Khodas}
\affiliation{CMPMSD, 
 Brookhaven National Laboratory, Upton, NY 11973 USA}
\affiliation{Physics Department,
 Brookhaven National Laboratory, Upton, NY 11973 USA}

%

\author{T.~Valla}
\author{I.~A.~Zaliznyak}
\affiliation{CMPMSD, 
 Brookhaven National Laboratory, Upton, NY 11973 USA}

\begin{abstract}

The magnetotransport in single layer graphene has been experimentally investigated in magnetic fields up to 18 T as a function of temperature. A pronounced T-dependence is observed for T $\lesssim 50$ K, which is either metallic, or insulating, depending on the filling factor $\nu$. The metal-insulator transition (MIT) occurs at $|\nu_c| \sim 0.65$ and in the regime of the dissipative transport, where the longitudinal resistance $R_{xx} > \frac{1}{2}R_{K}$. The critical resistivity ($R_{xx}$ per square) is $\rho_{xx}(\nu_c) \approx \frac{1}{2}R_{K}$ and is correlated with the appearance of zero plateau in Hall conductivity $\sigma_{xy}(\nu)$ and peaks in $\sigma_{xx}(\nu)$. This leads us to construct a universal low-T ($n$, B) phase diagram of this quantum phase transition.

\end{abstract}


\maketitle



Graphene, a honeycomb crystalline single layer of carbon atoms, is a gapless two-dimensional (2D) semimetal with many novel electronic properties \cite{Novoselov2004,Novoselov2005,Zhang2005,CastroNeto2009}. The low-energy charge excitations in graphene can be described by the Dirac equation for massless chiral fermions. In a perpendicular magnetic field B the electronic energy eigenstates become quantized Landau levels (LL) whose energies are given by $E_{N}=sgn(N)\hbar\omega_{c}\sqrt{\mid N\mid}$,
where $\omega_{c}=v_{F}\sqrt{2eB/(\hbar c)}$ is a ``cyclotron frequency'', $v_{F}$ is the fermion velocity and $N = 0,\pm1,\pm2,\pm3...$ is an integer. Unlike in a non-relativistic 2D electron gas (2DEG) system, there is a field-independent LL at $E=0$ for $N=0$ in graphene, which has a four-fold degeneracy, corresponding to two Fermi points of the honeycomb lattice ($K$ and $K'$ valleys), and two spin states. The $N = 0$ level is a fundamental property of the relativistic Dirac Hamiltonian and is particle-hole symmetric. As a result, an anomalous quantum Hall effect (QHE) has been observed in graphene, whose quantized Hall resistance is  given by $R_{xy} = \frac{1}{4} R_{K}/(N+\frac{1}{2})$, where $R_{K}=h/e^{2}$ is the resistance quantum \cite{Novoselov2005,Zhang2005}. A plateau at $R_{xy}=R_{K}/2$ appears at a filling factor $\nu=n_{s}h/(eB) = \pm 2$, where $n_{s}$ is the charge density per unit area, because only two of the four $N = 0$ LL are available to either electrons, or holes. In the semiclassical description of magnetotransport this leads to a Berry phase $\pi$, characteristic of Dirac particles.

\begin{figure}[!t]
\vspace{-0.25in}
\includegraphics[width=0.67\linewidth]{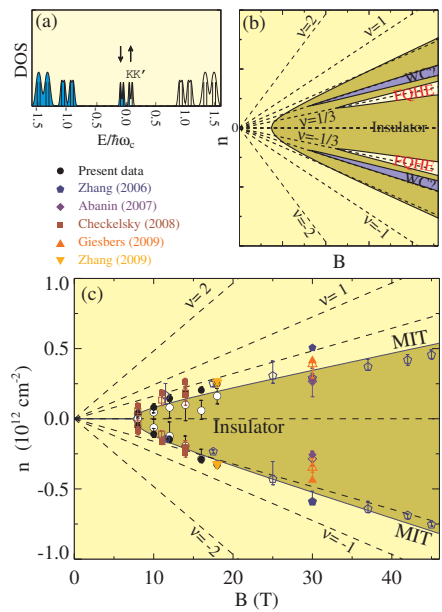}
\caption{(a) Splitting of the Landau levels in graphene. (b) Our proposed schematic phase diagram of the electronic states on the $N = 0$ LL consistent with the results of Refs. \cite{Zhang2006} - \cite{Feldman2009}. (c) The experimental phase diagram of the Quantum Hall Metal-Insulator transition in graphene obtained from our data. It also accommodates various data published in Refs. \cite{Zhang2006} - \cite{Feldman2009}, reconciling existing observations. Symbols are explained in the text.
 }
\label{PhaseDiagram}
\vspace{-0.25in}
\end{figure}

Electronic states with different spin and valley indices split even further (see Fig. \ref{PhaseDiagram} (a)). Spin-splitting results from the Zeeman interaction, and $K$ and $K'$ valley splitting from lattice imperfections and inter-valley scattering inherent to the mapping of the lattice Hamiltonian of graphene onto axially symmetric Dirac fermions \cite{Lukyanchuk_PRL2008}. The nature of the electronic states at the $N = 0$ LL has recently been under intense theoretical and experimental scrutiny, but it still remains unclear. While some theories predict unusual metallic transport via gapless edge states \cite{Fertig_PRL2006,AbaninLeeLevitov_PRL2006,Abanin2007} other suggest that a gap may open at high magnetic fields, resulting in an insulating state near the charge-neutrality point (CNP), where $n_{s} = 0$ \cite{Shimshoni2009,Nomura2009,DasSarma2009}.

Several groups have studied experimentally charge transport in the $N = 0$ LL regime near the CNP in graphene, with conflicting results \cite{Novoselov2005,Zhang2005,Abanin2007,Zhang2006,Jiang2007,Giesbers2007,Checkelsky2008,Giesbers2009,ZhangZaliznyak2009}.
Some groups have observed finite longitudinal resistance $R_{xx} \lesssim R_{K}$ even at high magnetic fields, where the QHE plateau at $R_{xy}=R_{K}/2$ is well developed \cite{Novoselov2005,Zhang2005,Abanin2007,Giesbers2007,Jiang2007}.
In contrast, others have found a marked resistance increase, to $R_{xx}\gg R_{K}$, and a plateau of nearly zero Hall conductivity near the CNP in high fields, suggesting a breakdown of the $N = 0$ QHE and an insulating state at low temperature \cite{Zhang2006,Checkelsky2008,Giesbers2009,ZhangZaliznyak2009}.

In some studies a plateau at $\nu = 1$ and $R_{xy}=R_{K}$ was also observed, indicating inter-valley splitting of the $N = 0$ LL, \cite{Zhang2006}. In Ref. \cite{Checkelsky2008}, $R_{xx}(B) \gtrsim 100 R_K$ was reported at the CNP, and its divergence with field was associated with a Kosterlitz-Thouless (KT) transition at a sample-dependent critical field $B_{c}$ ($ \gtrsim 18$ T), which was lower for better quality samples. The field-induced insulator behavior was later correlated with the filling factor $\nu$, and evidence for two different insulating phases emerging with decreasing $\nu \lesssim 1$ was found \cite{ZhangZaliznyak2009}. More recently, magnetic field driven insulator phases have been observed in both single and double layer high mobility suspended graphene samples, where they emerge from $\nu = 1$, as well as fractional, $\nu = 1/3$ QHE states \cite{Du2009,Bolotin2009,Feldman2009}.

%
%

The phase diagram of the electronic states at the $N = 0$ LL in graphene emerging from the above observations is quite complex. However, it seems to have a universal constituent, a metal-insulator transition (MIT) where the incompressible electronic liquid state of the QHE plateau is followed by a localized insulating phase [Fig. \ref{PhaseDiagram} (b)]. How does dissipative regime develop in a system of Dirac electrons in graphene, where robust quantum Hall transport is sustained up to hunderd K and above, is a fundamental open issue, which we address here by performing temperature-dependent magnetotransport measurements. We identify the first of these MIT, which occurs at $\nu \approx 0.65$, and establish its low-temperature ($n_{s}$, B) phase diagram, shown in Fig. \ref{PhaseDiagram} (c).


We have studied graphene samples prepared by mechanical exfoliation of pyrolytic graphite \cite{Novoselov2004} deposited onto a highly doped silicon substrate covered with 285 nm thick SiO$_2$ layer. Hall bar devices with Cr(3 nm)/Au(50 nm) contacts were patterned and etched using electron beam lithography and oxygen gas plasma etching, with the Si substrate serving as a gate to control the nature (electrons or holes) and density of the electronic carriers. (The optical image of the device investigated in this study is shown Fig. \ref{Fig1:SdH} (a)). Prior to measurements, devices were annealed for several hours in vacuum at 400 K, to remove impurities. Four-probe magnetoresistance measurements were performed using a low frequency ($7$ Hz) lock-in technique with a $I=10$ nA current.

%
\begin{figure}[!t]
\vspace{-0.25in}
\includegraphics[width=.67\linewidth]{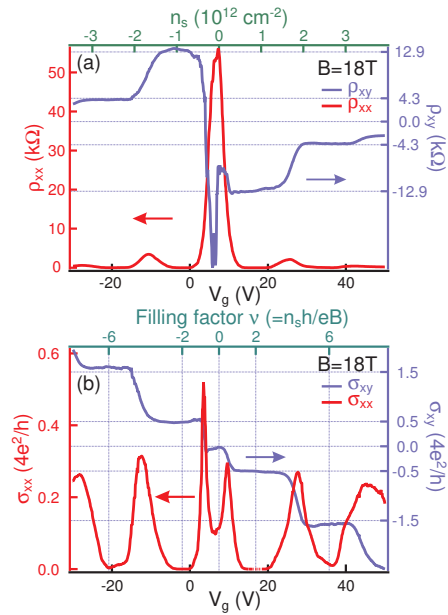}
\caption{QHE data as a function of the gate voltage $V_g$, for B = 18 T at T = 0.25 K, used in Fig.~\ref{Fig1:SdH}. (a) $\rho_{xx}$ and $\rho_{xy}$, top axis shows the corresponding carrier density $n_s$; (b) $\sigma_{xx}$ and $\sigma_{xy}$, top axis shows LL filling factor $\nu$. }
\label{Fig2:QHE}
\vspace{-0.25in}
\end{figure}
%


The longitudinal and Hall resistivities $\rho_{xx(xy)}$ \cite{R_Note} of our graphene sample at the base temperature (T = 0.25 K) and B = 18 T are shown in Fig.~\ref{Fig2:QHE}. The corresponding conductivities were obtained by standard matrix inversion, $\sigma_{xx(yx)}=\rho_{xx(xy)}/(\rho_{xx}^{2}+\rho_{xy}^{2}$). A peculiar behavior is observed near N = 0 LL filling, where $\rho_{xx}$ (and $R_{xx} \approx 3\rho_{xx}$) become large, noticeably exceeding $R_K$, and a two-peak structure in $\sigma_{xx}$ concomitant with a ``zero plateau'' of $\sigma_{xy} \simeq 0$ appears [Fig.~\ref{Fig2:QHE} (b)]. These observations agree with previously reported findings \cite{Abanin2007,Checkelsky2008,Giesbers2009,ZhangZaliznyak2009}.

The full ($n_{s}$, B) dependence of the longitudinal resistivity and conductivity at T = 0.25 K is summarized in Fig.~\ref{Fig1:SdH} in the form of color contour plots. The data was collected by gate voltage sweeps at fixed B every 2 T. To avoid errors from contacts misalignment, $\rho_{xy}$ and $\rho_{xx}$ were measured for two opposite directions of magnetic field and then antisymmetrized and symmetrized, respectively. As seen in Fig.~\ref{Fig1:SdH} (a), $\rho_{xx}$ displays well-developed Shubnikov\textendash{}de Haas (SdH) oscillations fanning away from the CNP, $V_0 \approx 7.5$ V, with increasing B. Maxima of $\rho_{xx}$ track LL filling $\nu = 0, \pm 2, \pm 6, \pm 10, ...$, which is typical of graphene \cite{Novoselov2005,Zhang2005,CastroNeto2009}. Despite the moderate mobility of our sample, ($\mu \approx 5.8\cdot 10^{3}$ cm$^{2}$/Vs), the QHE regime develops for B $\gtrsim 4$ T, with plateaus at $\rho_{xy} \approx \frac{1}{2} R_{K}$ and $\rho_{xx} \approx 0$.

A striking feature in Fig.~\ref{Fig1:SdH} (a) is a well-defined resistive region near $n_s = 0$ that appears at B $\gtrsim $ 9 T, where $\rho_{xx} \gtrsim R_K/2$. Its boundary is clearly correlated with maxima in $\sigma_{xx}$($n_s$, B) [Fig.~\ref{Fig1:SdH} (b)], and in the derivative of the Hall conductivity, $d\sigma_{xy}$($n_S$, B)$/d n_s$ [Fig.~\ref{Fig1:SdH} (c)]. The latter implies step-like behavior of $\sigma_{xy}$, corresponding to a plateau phase near $n_s = 0$.
While these observations clearly identify a well-defined region in ($n_s$, B) space corresponding to a resistive plateau phase near $n_s = 0$ for B $\gtrsim 9$ T, its nature remains unclear. Is this indeed a new electronic phase (insulator?) at the $N = 0$ LL, bounded by a phase-transition line in the ($n_s$, B) phase diagram?

%
\begin{figure}[!b]
\vspace{-0.2in}
\includegraphics[width=.8\linewidth]{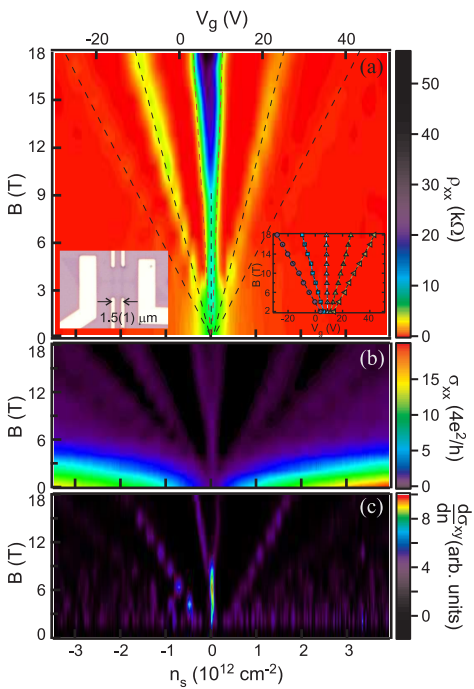}
\caption{(a) Contour map of the longitudinal resistivity, $\rho_{xx} = R_{xx} \cdot W_s/L_s$, as a function of charge density $n_s$ and magnetic field B, $L_s = 1.5(1) \mu$m, $W_s = 0.5(1) \mu$m are sample dimensions. Top axis shows the back gate voltage $V_{g}$. LL are clearly identified by ridges fanning away from $n_s = 0$ with increasing $B$, traced by black broken lines. Left bottom inset is the optical image of the device. Right bottom inset shows the LL fan diagram with fits yielding $n_s = C_V(V_g-V_0)$, $C_V = 9.2(4) \cdot 10^{10}$ cm$^{-2}$/V, for charge density and $V_0 = 7.5(1)$ V for the charge neutrality point (CNP). (b) The corresponding maps of the longitudinal conductivity, $\sigma_{xx}$, and (c) the derivative of the Hall conductivity, $d\sigma_{xy}/dn_s$.
 }
\label{Fig1:SdH}
\vspace{-0.2in}
\end{figure}
%

\begin{figure}[!t]
\vspace{-0.25in}
\includegraphics[width=.8\linewidth]{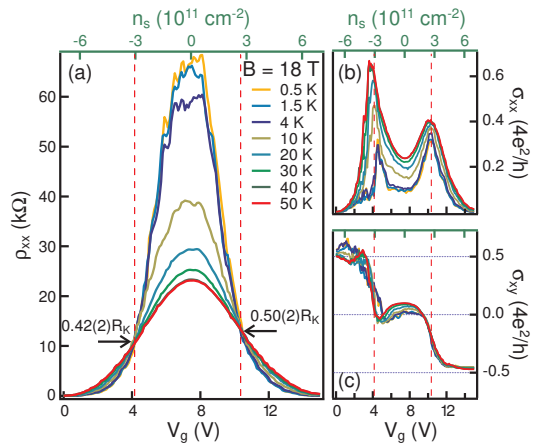}
\caption{Temperature dependence of the longitudinal resistivity (a), and longitudinal (b), and Hall (c), conductivities versus $V_{g}$ and carrier density (top axes) at B = 18 T. Vertical broken lines show critical densities separating the insulating and the metallic phases, and are determined from the crossing points in panel (a).
 }
\label{Fig3:RxxSxxSxy_Tdep}
\vspace{-0.25in}
\end{figure}

In order to answer this question and understand the appearance of the resistive $N=0$ LL state in Fig.~\ref{Fig1:SdH}, we studied the temperature dependence of the magnetoresistance data at B = 18 T shown in Fig.~\ref{Fig2:QHE}. The results of gate voltage sweeps at T = 0.5, 1.5, 4, 10, 20, 30 and 50 K are shown in Fig. \ref{Fig3:RxxSxxSxy_Tdep}. Two regions are easily identified in the figure: a low-$n_s$ region where the longitudinal resistance increases with decreasing temperature, and another region at high carrier densities in which the resistance shows the opposite, metallic behavior [Fig. \ref{Fig3:RxxSxxSxy_Tdep} (a)]. An obvious feature separating the two regions are crossing points of the curves measured at different temperatures, where $\rho_{xx}$ is temperature-independent. Inspection of Fig. \ref{Fig3:RxxSxxSxy_Tdep} (b), (c) shows that the insulator phase at low $n_s$ is marked by the appearance of a $\sigma_{xy} \simeq 0$ plateau and an accompanying two-peak structure in $\sigma_{xx}$. Thus, a low-$n_s$ plateau phase identified by the behavior of the conductivity in Figures \ref{Fig2:QHE}, \ref{Fig1:SdH} at T = 0.25 K for B $\gtrsim 9$ T, indeed corresponds to a distinct, insulating phase.

%
%

Metal-insulator transition in Fig. \ref{Fig3:RxxSxxSxy_Tdep} occurs for $\nu_c^{(e)} \approx 0.57$, $\rho_{xx}^{(e)} \approx 0.5 R_K$ and $\nu_c^{(h)} \approx -0.75$, $\rho_{xx}^{(h)} \approx 0.42 R_K$, showing some 20\% asymmetry between the electron and hole sides, respectively. However, the critical resistivity is remarkably close to the value $R_K/2$ expected for a metal-insulator transition in disordered 2D electron systems (the denominator 2 is a consequence of graphene valley degeneracy) \cite{Punnoose2005}.
Since contacts in our etched Hall bar device are non-invasive, the observed electron-hole asymmetry is unlikely to be contact-related. Perhaps, it arises from breaking of the particle-hole symmetry by random external potential, suggesting a connection of the observed MIT with electrostatic disorder.

A similar behavior of resistivity isotherms measured at different T as a function of magnetic field has previously been observed in conventional 2DEG systems at low carrier densities \cite{Jiang1993,Hughes1994,Shahar1995,Pruisken2000,Dolgopolov1992}.
The crossing point of the $\rho_{xx}(B, T)$ curves where the sign of the temperature derivative $d\rho_{xx}/dT$ changes, has been associated with a transition from the electronic liquid state of the QHE plateau phase to an insulator, as the LL filling is depleted with increasing B. Such plateau-insulator transitions (PIT) have been observed both for integer and fractional QHE plateaus and are understood as metal-insulator quantum phase transition induced by Anderson (strong) localization of electrons. In conventional 2DEGs the transition from the last integer QHE plateau to an insulator occurs at $\rho_{xx} \sim R_K$ and in the range $0.52 \lesssim \nu_c \lesssim 0.8$, similar to what we find in graphene, even though massless charge carriers are expected to avoid strong localization and the localization mechanism in graphene must be of different nature.

\begin{figure}[!t]
\includegraphics[width=.9\linewidth]{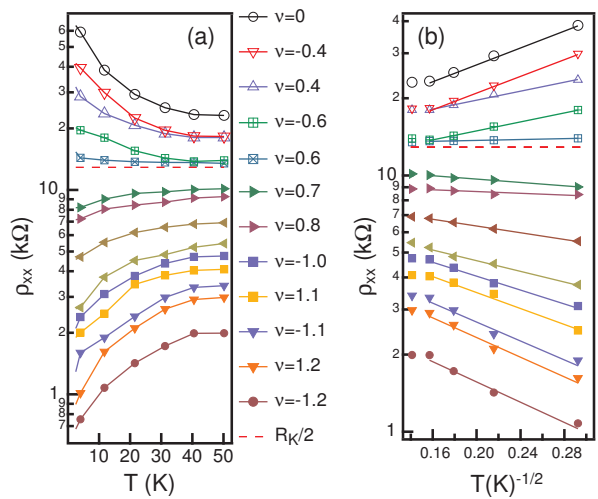}
\caption{(a) Semi-log plot of $\rho_{xx}$ versus temperature at $B=18$ T for 12 filling factors $\nu$, revealing the metal-insulator transition at $\nu_c \approx -0.7$ and $0.6$, where $\rho_{xx}{(\nu_c)} \approx R_K/2$ (dashed line). (b) Same data as a function of $T^{-1/2}$.
 }
\label{Fig4:Rxx_T}
\vspace{-0.25in}
\end{figure}

Association of the MIT with the crossing point of resistivity isotherms is only meaningful when $\rho_{xx}(B, T)$ shows clear metallic and insulating behavior in a finite temperature range on the two sides of the transition \cite{Dolgopolov1992}. As shown in Fig. \ref{Fig4:Rxx_T}, this is indeed the case for 1 K $\lesssim$ T $\lesssim 50$ K for the B = 18 T data in our graphene sample. In a temperature window 10 K $\lesssim T \lesssim$ 50 K, $\log \rho_{xx}(B, T)$ is consistent with nearly linear dependence on $T^{-1/2}$ [Fig. \ref{Fig4:Rxx_T} (b)], indicative of Coulomb-gap-induced variable range hopping (VRH). Despite the limited dynamical range of our data, such dependence agrees visibly better than a simple activation dependence $T^{-1}$. This provides a hint that scattering mechanism driving the insulator state is a screened Coulomb interaction.
Remarkably, the same T-dependence is retained across the transition, persisting with the slope sign change on the metallic side. Similar symmetry, where $\rho_{xx}(T, \nu - \nu_c) \propto 1/\rho_{xx}(T, \nu_c - \nu)$, has also been observed in low mobility 2DEG \cite{Shahar1995}.

%
%

Having established the signature of the MIT on the $N=0$ LL in graphene, we now return to constructing its low-T (B, $n_s$) phase diagram. This is shown in Fig. \ref{PhaseDiagram} (c), which complies various data reported earlier by other groups along our own results. Closed symbols show transition identified from the longitudinal resistivity, using $\rho_{xx} = R_K/2$, while open symbols correspond to the $\sigma_{xy} \simeq 0$ plateau phase, identified from $\sigma_{xy}(\nu_c) = 1/(2R_K)$, i. e. the mid-point between the $N = 0$ and $N = 1$ plateaus. The overall topology of the phase diagram, including the electron-hole asymmetry, is remarkably universal, but the exact MIT critical fields are slightly sample-dependent, especially at low B.


In conclusion, we have presented evidence for a metal-insulator 
transition on the $N = 0$ LL in graphene, which occurs in the regime of the dissipative transport, where $R_{xx} > R_K/2$, and at $\rho_{xx} \approx R_K/2$ ($R_{xx} \approx 3\rho_{xx}$ for our sample). %
It is surprisingly similar to the plateau-insulator transition in 2DEG \cite{Jiang1993,Hughes1994,Shahar1995,Pruisken2000,Dolgopolov1992}, even though strong localization driving the PIT in 2DEGs should be absent in graphene. This difference is clearly seen in the low-field part of the phase diagram of Fig. \ref{PhaseDiagram}, where Klein tunneling of Dirac electrons in graphene impedes the low-density insulating phases \cite{Dolgopolov1992}.
In 2DEGs, the conductivities in the vicinity of the PIT follow a semicircle relation \cite{Dykhne1994}, indicating a transition to a peculiar quantum Hall insulator state \cite{Shahar1995,Kivelson1992}. It was found that in graphene a semicircle describes very well the transitions between plateaus in the metallic phase \cite{Abanin2007,Burgess2007}. For the $N = 0$ LL it predicts $\rho_{xx}^2 + \rho_{xy}^2 = (R_K/2)^2$, and therefore $\rho_{xx} \leq R_K/2$. We find that this is violated in the insulating phase, suggesting that $\rho_{xx} \approx R_K/2$ ($\sigma_{xx} \approx 2e^2/\hbar$) is the maximum metallic resistivity (minimum conductivity) of the dissipative Hall state on the $N = 0$ LL in graphene.

\begin{acknowledgments}

We thank E. Mendez for insightful discussions and valuable comments on the paper. Material preparation and device processing was done at Brookhaven National Laboratory's Center for Functional Nanomaterials. This work was supported by the US DOE under Contract DE-\-AC02-\-98CH10886. The work of Yan Zhang is supported by NSF DMR-0705131. Magnetic field experiments were carried out at NHMFL, which is supported by NSF through DMR-0084173 and by the State of Florida.

\end{acknowledgments}


\begin{thebibliography}{10}
\bibitem{R_Note}
Resistivity, $\rho_{xx} = R_{xx} \cdot W_s/L_s$ where $W_s$ and $L_s$ are sample width and length, is a proper characteristic of the dissipative transport, while in the regime of quantum Hall transport, the resistance $R_{xx} \lesssim R_{K}$ is fundamental.
\end{thebibliography}

\end{document}